\documentclass[pdflatex,sn-basic]{sn-jnl} 
\jyear{2022}%

\theoremstyle{thmstyleone}%
\theoremstyle{thmstyletwo}%

\theoremstyle{thmstylethree}%

\newcommand{\bm}[1]{{\mbox{\boldmath $#1$}}}
\usepackage{aas_macros}
\begin{document}

\title[Dynamics of large-scale solar flows]{Dynamics of large-scale solar flows}


\author[1]{\fnm{Hideyuki}~\sur{Hotta}}\email{hotta.h@isee.nagoya-u.ac.jp}
\author[2]{\fnm{Yuto}~\sur{Bekki}}\email{bekki@mps.mpg.de}
\author*[2,3]{\fnm{Laurent}~\sur{Gizon}}\email{gizon@mps.mpg.de}
\author[4,5] {\fnm{Quentin}~\sur{Noraz}}\email{quentin.noraz@astro.uio.no}
\author[6] {\fnm{Mark}~\sur{Rast}}\email{mark.rast@colorado.edu}

\affil[1]{\orgdiv{Institute for Space-Earth Environmental Research}, \orgname{Nagoya University}, \orgaddress{\city{Chikusa-ku, Nagoya, Aichi}, \postcode{464-8601}, \country{Japan}}}

\affil[2]{\orgname{Max-Planck-Institut f\"{u}r Sonnensystemforschung}, \orgaddress{\street{Justus-von-Liebig-Weg 3}, \city{G\"{o}ttingen}, \postcode{37077}, \country{Germany}}}

\affil[3]{\orgdiv{Institut für Astrophysik}, \orgname{Georg-August-Universt\"{a}t G\"{o}ttingen}, \orgaddress{\street{Friedrich-Hund-Platz 1}, \city{G\"{o}ttingen}, \postcode{37077}, \country{Germany}}}

\affil[4]{\orgdiv{Rosseland Centre for Solar Physics}, \orgname{University of Oslo}, \orgaddress{\street{P.O. Box 1029 Blindern}, \city{Oslo}, \postcode{NO-0315}, \country{Norway}}}

\affil[5]{\orgdiv{Institute of Theoretical Astrophysics}, \orgname{University of Oslo}, \orgaddress{\street{P.O. Box 1029 Blindern}, \city{Oslo}, \postcode{NO-0315}, \country{Norway}}}

\affil[6]{\orgdiv{Department of Astrophysical and Planetary Sciences, Laboratory for Atmospheric and Space Physics}, \orgname{University of Colorado}, \orgaddress{\city{Boulder}, \state{CO} \postcode{80309}, \country{USA}}}


\abstract{
The Sun's axisymmetric large-scale flows, differential rotation and meridional circulation, are thought to be maintained by the influence of rotation on the thermal-convective motions in the solar convection zone. These large-scale flows are crucial for maintaining the Sun's global magnetic field. 
Over the last several decades, our understanding of large-scale motions in the Sun has significantly improved, both through observational and theoretical efforts. Helioseismology has constrained the flow topology in the solar interior, and the growth of supercomputers has enabled simulations that can self-consistently generate large-scale flows in rotating spherical convective shells. In this chapter, we review our current understanding of solar convection and the large-scale flows present in the Sun, including those associated with the recently discovered inertial modes of oscillation. We discuss some issues still outstanding, and provide an outline of future efforts needed to address these.
}




\maketitle

\newpage 

\tableofcontents

\pagebreak

\section{Observations of large-scale flows in the Sun}
\label{sec:helioseismology}

\subsection{Solar differential rotation and meridional circulation}\label{sec:solDRMC}

\vspace{0.35cm}
\leftline{{{\bf Differential rotation}}}
\vspace{0.35cm}

Due to the solar rotation, axisymmetry (about its rotational axis) of the large-scale flows is established to some degree in the Sun's interior.
\textit{Differential rotation} denotes the longitudinal component of the axisymmetric (longitudinally-averaged) flow which varies with radius and latitude.
It arises from the nonlinear interaction of the rotationally-influenced solar magneto-convection \citep[e.g.,][]{miesch2005}.
Differential rotation represents a shear in the rotation rate and is thought to play a significant role in the solar dynamo by stretching and amplifying the magnetic field lines \citep[][]{charbonneau2020}.

\begin{figure}[t]
\centering
\includegraphics[width=0.99\textwidth]{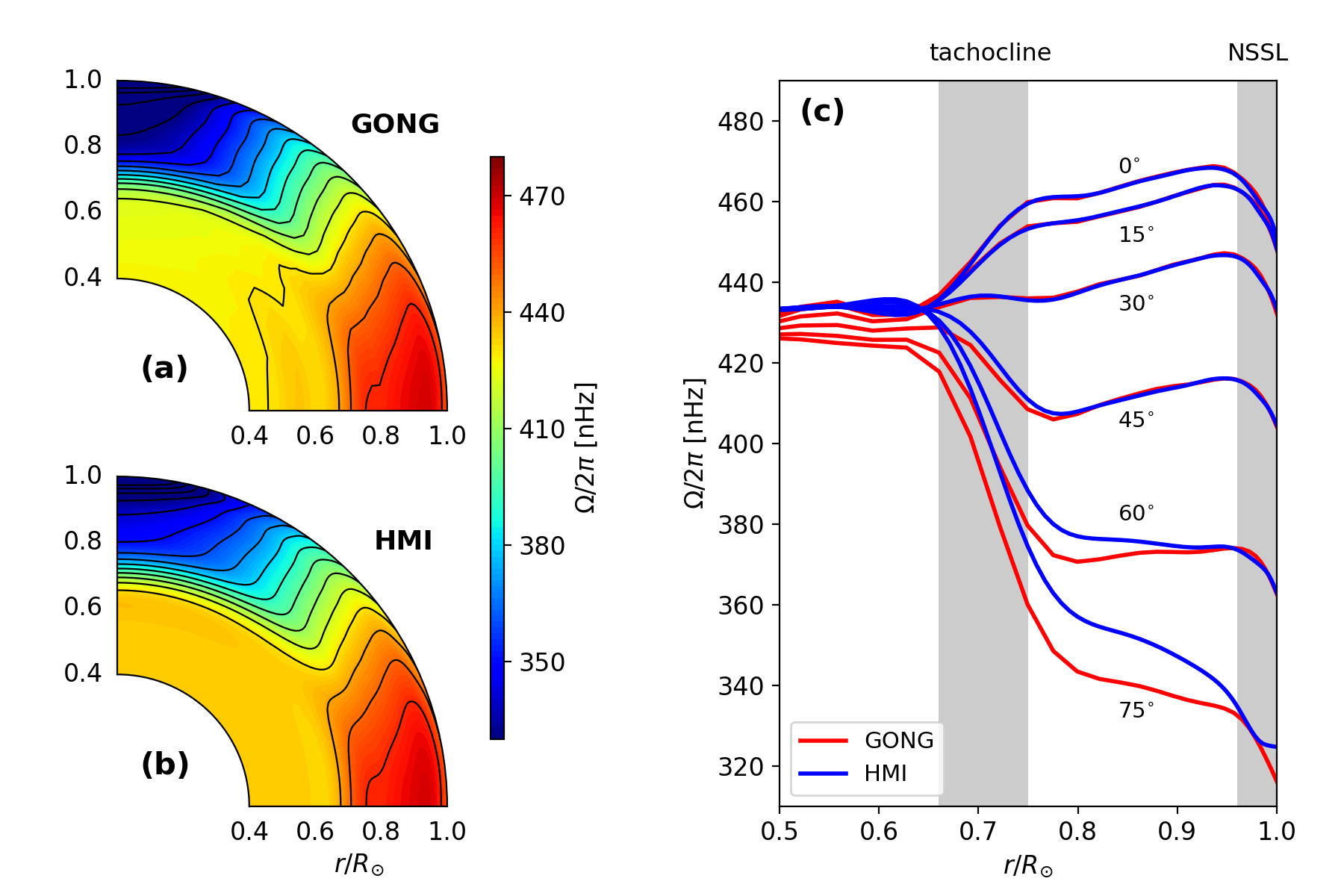}
\caption{
Internal profile of the solar differential rotation deduced from global helioseismology and averaged from April 2010 to February 2021. Panels (a) and (b) show
the results obtained by the Global Oscillation Network Group (GONG) \citep[data courtesy of R.~Howe, using the method of][]{howe2005} and the Helioseismic and Magnetic Imager (HMI) onboard the Solar Dynamics Observatory (SDO) \citep[][]{larson2018}.
Panel (c) shows the radial differential rotation at selected latitudes.
Grey shades denote the layers of strong radial rotational shear known as the tachocline and the near-surface shear layer (NSSL). 
}\label{fig_bekki:dr}
\end{figure}

The solar differential rotation profile can be measured by \textit{global helioseismology}, which analyzes small frequency splittings of resonant acoustic oscillations (global standing acoustic modes) \citep[][]{duvall1984,thompson1996,schou1998,howe2000}.
Figure~\ref{fig_bekki:dr} shows the observationally-inferred profile of the internal differential rotation of the Sun \citep[][]{howe2005,larson2018}.
We summarize striking features of the solar differential rotation as follows:
\begin{itemize}
    \item The radiative interior rotates almost rigidly.
    \item In the convection zone, the equator rotates about $30\%$ faster than the poles.
    \item The transition from uniformly-rotating radiation zone to differentially-rotating convection zone occurs in a thin layer from $0.68R_{\odot}$ to $0.73R_{\odot}$.
    This layer is called the \textit{tachocline}.
    \item In the bulk of the convection zone ($0.73R_{\odot} < r < 0.96R_{\odot}$), the rotation rate varies strongly with latitude and much more weakly with radius.
    \item In a shallow surface layer ($r \gtrsim 0.96R_{\odot}$), the rotation rate decreases by about $5\%$ at all latitudes. 
    This layer is called the \textit{near surface shear layer}.
    \item The contours of constant angular velocity are inclined by about $25^{\circ}$ with respect to the rotational axis over a wide range of latitude. In other words, the differential rotation does not follow the Taylor-Proudman theorem.
\end{itemize}
These observational facts need to be explained by theoretical and numerical models of rotating solar magneto-convection.

\vspace{0.7cm}
\leftline{{{\bf Meridional circulation}}}
\vspace{0.35cm}

\textit{Meridional circulation} represents radial and latitudinal components of the large-scale axisymmetric flow in the Sun, i.e., a poloidal flow in a meridional plane.
Meridional circulation, as well as the differential rotation, is believed to play a significant role in the solar dynamo by advecting the magnetic flux in both radial and latitudinal directions \citep[e.g.,][]{charbonneau2020}.

The meridional circulation is much weaker than the differential rotation (two orders of magnitudes smaller in flow amplitude) and cannot be inferred using conventional global-mode helioseismology.
Therefore, it is an extremely difficult task to measure the meridional flow in the Sun.
Near the solar surface, the meridional flow is poleward in both hemispheres with typical amplitudes of $\sim 10 -20$ m~s$^{-1}$.  This was first measured by \citet[][]{duvall1979} using Doppler measurements and then robustly confirmed in follow-up studies  by a variety of methods \citep[e.g.,][]{patron1995,giles1997,hathaway1996,braun1998,haber2002,ulrich2010,basu2010}.

Local helioseismology can extend these measurements into the deeper convection zone \citep[e.g.,][]{gizon2005}.
In particular, time-distance helioseismology \citep{duvall1993} and ring-diagram analysis \citep{hill1988} can be used to measure the effects of the meridional flow on north-south propagating waves. In time-distance helioseismology, wave travel times are extracted from the cross-covariance of the Doppler signals between points along meridians. The technique was applied first by \citep{giles1997} to the Michelson Doppler Imager (MDI) data onboard SOHO \citep[][]{scherrer1995}. However, this is an extremely difficult measurement to make because the deep meridional circulation is very weak (no greater than $3-5$ m~s$^{-1}$) and the sensitivity of the travel times to flows also decreases with depth. The analysis of very long time series is required to reduce noise. Furthermore, for accurate measurements it is critically important to apply corrections to the measurements, especially a center-to-limb correction \citep[][]{duvall2009,zhao2012}, and corrections for the $P$-angle and the Carrington elements  \citep[e.g.][]{giles2000, hathaway2010,liang2017}.
Often pixels that are in regions of very strong magnetic fields (e.g. in sunspots) are excluded from the cross-covariances \citep[e.g.][]{liang2015}.
The center-to-limb effect is, in general, very significant and depends strongly on time and instrument \citep{Liang2018}. It may have both an instrumental and physical component. When \citet{giles1997} made their measurements, no center-to-limb correction was applied to the MDI travel times, as it was very small during the year 1996 \citep{Liang2018}. We refer the reader to the review by  \citet{hanasoge2022} for additional information.

\begin{figure}[t]
\centering
\includegraphics[width=0.85\textwidth]{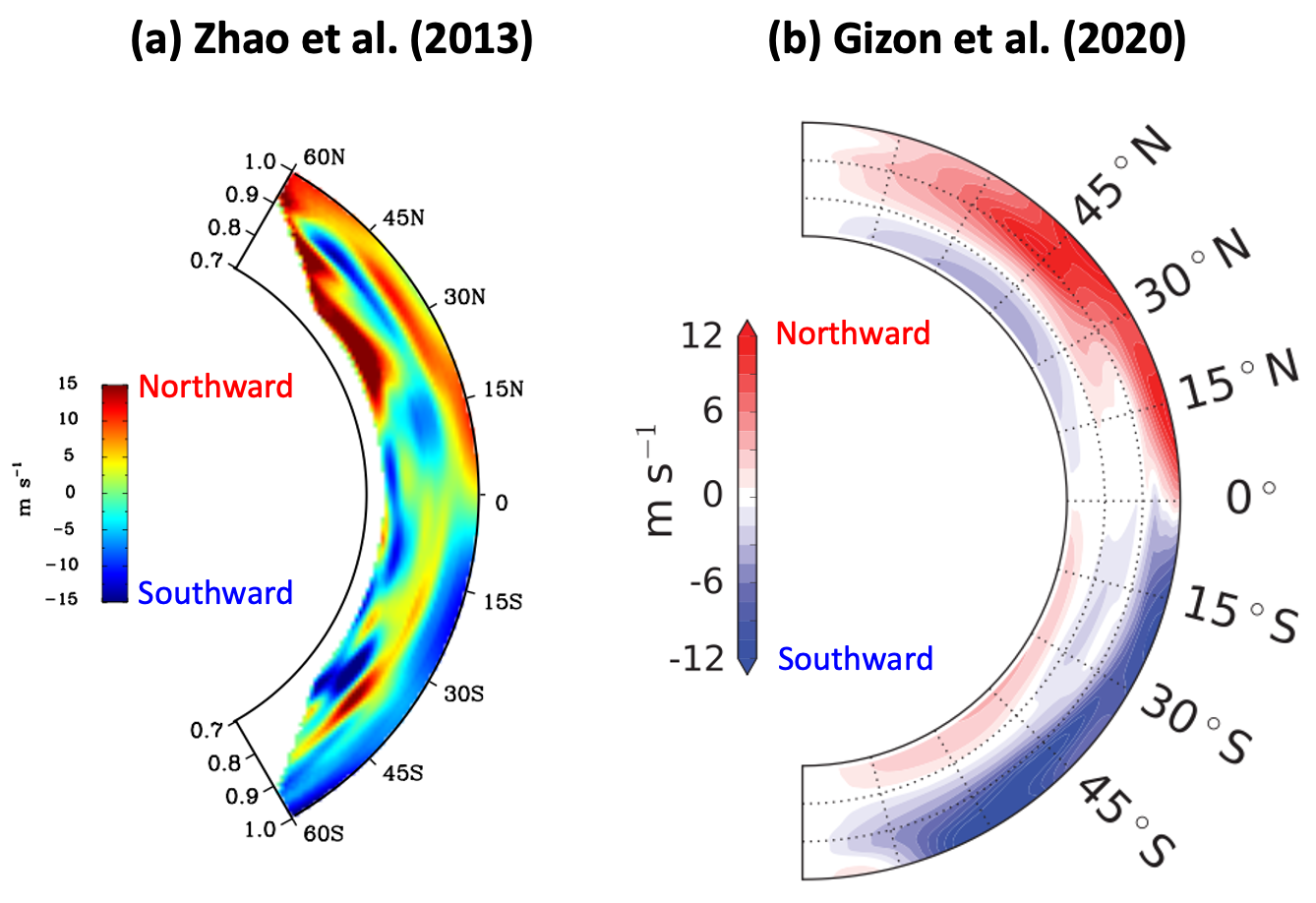}
\caption{
Latitudinal component of the meridional flow inferred by time-distance local helioseismology.
The red and blue shades correspond to the northward and southward directions respectively.
(a) The result obtained by \citet[][]{zhao2013} using SDO/HMI data (2010--2012), reproduced with permission.
(b) The result obtained by \citet[][]{gizon2020s} using GONG data (2008--2019), reprinted with permission from AAAS.
}\label{fig_bekki:mc}
\end{figure}

Many different inferences of the solar meridional circulation using time-distance helioseismology have been published. They are not consistent. Using travel time measurements obtained from SDO/HMI in 2010-2012, \citet[][]{zhao2013} reported an equatorward return flow in the middle of the convection zone ($0.82$--$0.91R_{\odot}$) and a poleward flow below $0.82R_{\odot}$, indicating a double-cell structure in radius (Fig.~\ref{fig_bekki:mc}a).
A similar result was obtained by \citet[][]{chen2017} who used seven years of data from HMI and active regions were masked out to remove the contribution from pixels with strong magnetic fields.
Note, however, that \citet[][]{zhao2013} and \citet[][]{chen2017} did not invert the radial component of the meridional flow which is required by the local mass conservation.
An attempt to test a mass conservation constraint was made by \citet[][]{jackiewicz2015} who found that the equation of continuity is poorly satisfied for the inverted flows from GONG and HMI (only two years of data were used).
Under the constraint of mass conservation in terms of the stream function, \citet[][]{rajaguru2015}, using four years of HMI data report a single-cell meridional circulation in each hemisphere, with an equatorial return flow near the base of the convection zone (below $0.77R_{\odot}$). A similar result was reported by \citet[][]{mandal2018}.

\citet[][]{liang2017} have compared the north-south travel-time data from SOHO/MDI and those from SDO/HMI over the 1-year overlap period of 2010. After correcting for the center-to-limb effect \citep{zhao2012}, it is found that the travel-times are consistent within the error bars, although an overlap period of only 288 days is far from enough for such consistency test. When comparing the travel-times with different forward models of meridional flows, \citet{Liang2018} found that the MDI/Cycle 23 data point to a single cell meridional flow in both hemispheres, while the HMI/Cycle 24 data point to a double cell in the north and a single cell in the south.

Following this work, \citet[][]{gizon2020s} carried out comprehensive measurements of the north-south travel-time measurements using all available data sets from GONG (2001-2019), SOHO/MDI (1996-2011), and SDO/HMI (2010-2019). After correcting for all known systematic errors (CCD orientation, center-to-limb effect, Carrington elements, pixels in sunspots and active regions), it was found that the north-south travel times measured from GONG and MDI are in good agreement over the overlap period of 1996-2011. However, a small travel-time offset is apparent in the HMI data compared to the GONG data over the period 2010-2019. No explanation was found for this HMI offset and the data were set aside:
The HMI measurements imply a strikingly different meridional flow pattern between the northern and southern hemispheres, as hinted at by \citet{Liang2018} and confirmed by \citet{gizon2020s} and \citet{braun2021}. 
\citet{gizon2020s} find that the mass-conserving meridional flows during cycle 23 (from MDI/GONG data) and cycle 24 (from GONG data) display a  single-cell pattern in each hemisphere, as shown in Fig.~\ref{fig_bekki:mc}b. 

Work is needed to resolve the remaining issues that affect helioseismic inferences of the deep meridional flow. Efforts are ongoing to identify the source of the HMI travel-time offset, which is extremely small ($0.2$~s) but significant. The other pressing issue is to gain an understanding of the origin of the center-to-limb effect \citep[e.g.][]{chen2018}. The analysis of travel-times in different frequency bands is a promising avenue to make progress in this area \citep{chen2018,rajaguru2020}.

Observational determination of the solar meridional circulation is crucial, not only for constraining solar dynamo models \citep[e.g.,][]{hazra2014}, but also for properly understanding the angular momentum flux balance in the Sun's convection zone \citep[e.g.,][]{featherstone2015}.

\vspace{0.7cm}
\leftline{{{\bf Cycle variations of large-scale mean flows}}}
\vspace{0.35cm}

Solar Doppler observations and global helioseismology have revealed that that the Sun's differential rotation exhibits a temporal variation pattern consisting of multiple bands of faster- and slower-than-average zonal flows which migrate in latitude with the phase of the solar dynamo cycle \citep[e.g.,][]{howe2009}.
The so-called \textit{torsional oscillation} \citep[coined by][]{howard1980} has two distinct branches, one at active-region latitudes and one at latitudes above $50^\circ$.
The low-latitude branch migrates equatorward together with the magnetic activity belts and shows a clear $11$-yr periodicity.
On the other hand, the poleward-migrating high-latitude branch tends to show a rather irregular behavior: it was clearly seen in Cycle 23 but did not appear in Cycle 24 \citep[e.g.,][]{howe2013}.
For more details on the observational aspects of the Sun's torsional oscillations, see \citep[][section 6.2]{norton2023}.
The amplitude of the solar torsional oscillation is $\approx 3$--$5$~nHz and it is important to understand its physical origin as it likely reflects the nonlinear feedback of the dynamo-generated magnetic fields onto the large-scale flows \citep[e.g.,][]{rempel2007,beaudoin2013,pipin2019,brunPoweringStellarMagnetism2022}.

As is the case of differential rotation, the poleward meridional circulation shows a temporal variation associated with the solar magnetic cycle \citep{Beck2002,Gizon2004, Gizon2010,Mahajan2023}; see also Section 6.3 in Chapter 2 \citep[][]{norton2023}.
This variation is of order $\approx 5$~m~s$^{-1}$ or more, which is a significant fraction of the maximum meridional flow amplitude. It may be explained, at least in part, by the north-south component of the inflows around active regions \citep{Gizon2010, Mahajan2023}. 
In addition to this component, \citet{Mahajan2023} find that there is a residual solar-cycle component as small as $\approx 2$~m~s$^{-1}$, which is seen around cycle minima.

\subsection{Solar convective flows}
    
Convection on the Sun occurs over a wide range of spatial scales, and while the spectrum is continuous, apparent characteristic scales are commonly cited: granulation, mesogranulation, supergranulation, and giant cells.  Granulation~\citep{1801RSPT...91..265H} is readily apparent in high-resolution images of the solar photosphere, as a pattern of bright upflowing regions separated by darker downflowing lanes.  The characteristic upflow cells have diameters of $\sim1000~$km, lifetimes of about $0.2~$hr, and vertical flow speeds of $\sim1~$km~s$^{-1}$.  The upflow velocity often peaks near the granular boundaries~\citep[e.g.,][and reference therein]{1992A&A...253..561N, 1995ApJ...443..863R, 2002A&A...392.1105H, 2009LRSP....6....2N, 2017A&A...605A..87F}.  These properties reflect the compressible flow dynamics of a strongly cooled radiative boundary layer, with observations confirming the convective nature of the flow via measurement of the correlation between the vertical velocity and plasma temperature~\citep[e.g.,][]{ 1973SoPh...33...33C}.  Granulation is well observed and robustly modeled~\citep[se e.g.,][]{ 2009LRSP....6....2N}, even in quite shallow domains, by codes that capture the rapid change in radiative opacity in the solar photosphere and implement an open lower boundary condition to minimize bottom-up influences on the top-down dynamics of the radiative boundary layer.

Mesogranulation~\citep[][]{1981ApJ...245L.123N}, on the other hand, is observationally elusive.  With a reported length scale of about $5 -10~$Mm, $\sim60~$m~s$^{-1}$ vertical flow speeds, and $\sim2-3~$hr lifetime, its identification as a convective feature is still debated. Most recent studies suggest that no distinct mesogranular scale is present in the broad range of convective scales observed~\citep[e.g.,][and references therein]{2018LRSP...15....6R}.  One possibility is that there is weak advective self-organization of the granular flows, a process first proposed in the context of supergranulation~\citep[][]{2000A&A...357.1063R,  2003ApJ...597.1200R} but likely more relevant on mesogranular scales~\citep[][]{2001ApJ...563L..91C, 2005ApJ...632..677B, 2005A&A...444..245L, 2010ApJ...725L..47D}.  However, the absence of a mesogranular scale in the clustering of magnetic elements in high resolution magnetograms suggests that this mechanism too leads to a continuous exponential distribution of scales between 2 and $10~$Mm, with no distinctive  characteristic peak~\citep[][]{2013SoPh..282..379B}. 

Supergranulation~\citep[][]{1954MNRAS.114...17H, 1962ApJ...135..474L} is the largest likely-convective scale of motion readily visible in the solar photosphere.  It is observed directly in spectral Doppler shifts away from disk center (due to horizontal motions) and is traced by network magnetic elements which are prominent in magnetograms and in emission in low chromospheric lines such as Ca II K.  There is good correlation between Ca II K emission and magnetic flux density~\citep[][and references therein]{2005MmSAI..76.1018O}.  Supergranular cells have diameter of $\sim30~$Mm, horizontal flow velocities of $\sim100~$m~s$^{-1}$, and lifetimes of $\sim20~$hr.  After the intensity contributions of the small scale magnetic field elements has been removed, they show an average continuum intensity contrast across the cells of about $0.1\%$, corresponding to about one degree Kelvin in brightness temperature~\citep[][]{2009ApJ...707...67G}.  

The origin of the supergranular motions has been widely debated~\citep[see][and references therein]{ 2018LRSP...15....6R}.  It has recently been proposed that the scale of supergranulation reflects not a selected convective scale, but is instead defined by the scale above which convective power declines~\citep[][]{lord2014, 2016ApJ...829L..17C}. This interpretation, and the reasons underlying the power reduction, links the well observed phenomenon of supergranulation to the \textit{convective conundrum}, an outstanding discrepancy between models and observations (see this section below, and Sections~2.2 and 3.1). We note that the early suggestion that helium ionization plays a role in determining the mesogranular and supergranular scales~\citep[][]{1962ApJ...135..474L, 1964ApJ...140.1120S} is not supported by numerical simulations or simplified models base on them~\citep[][]{1993ApJ...419..224R, lord2014}.  Additionally, the presence of the network magnetic elements themselves~\citep[][]{2007ApJ...662..715C, 2012ApJ...757..187T} or the enhanced radiative loses through them~\citep[][]{2003ApJ...597.1200R} do not seem to play a role in scale selection role~\citep[][]{2014PhDT.......241L}.  Finally, it is important to note that supergranulation shows peculiar unexplained wave-like properties~\citep[][]{2003Natur.421...43G, 2003ApJ...596L.259S, 2018A&A...617A..97L}.    

In contrast with supergranulation, which is readily observed but not captured by any local-area or global spherical-shell simulation, solar giant cells~\citep[][]{1968ZA.....69..435S}, motions on the scale of the solar convection zone depth ($\sim 200~$Mm), dominate global spherical-shell simulations but are very difficult to observe~\citep[][and references therein]{2013Sci...342.1217H, 2021ApJ...908..160H}.  If, in the Sun, giant-cells had the amplitude they do in simulations, they would be easily observed in the solar photosphere.  This is the simplest manifestation of the convective conundrum:  that supergranulation, rather than giant-cell scale motions, are the largest readily observed motions in the solar photosphere.  The implication for solar differential rotation is fundamental.  
The enhanced amplitude of the  large-scale convective motions in global numerical simulations tend to place those simulations in a Rossby-number regime that favors anti-solar differential rotation profiles.

\begin{figure}[!t]
\centering
\includegraphics[width=9cm,height=5cm]{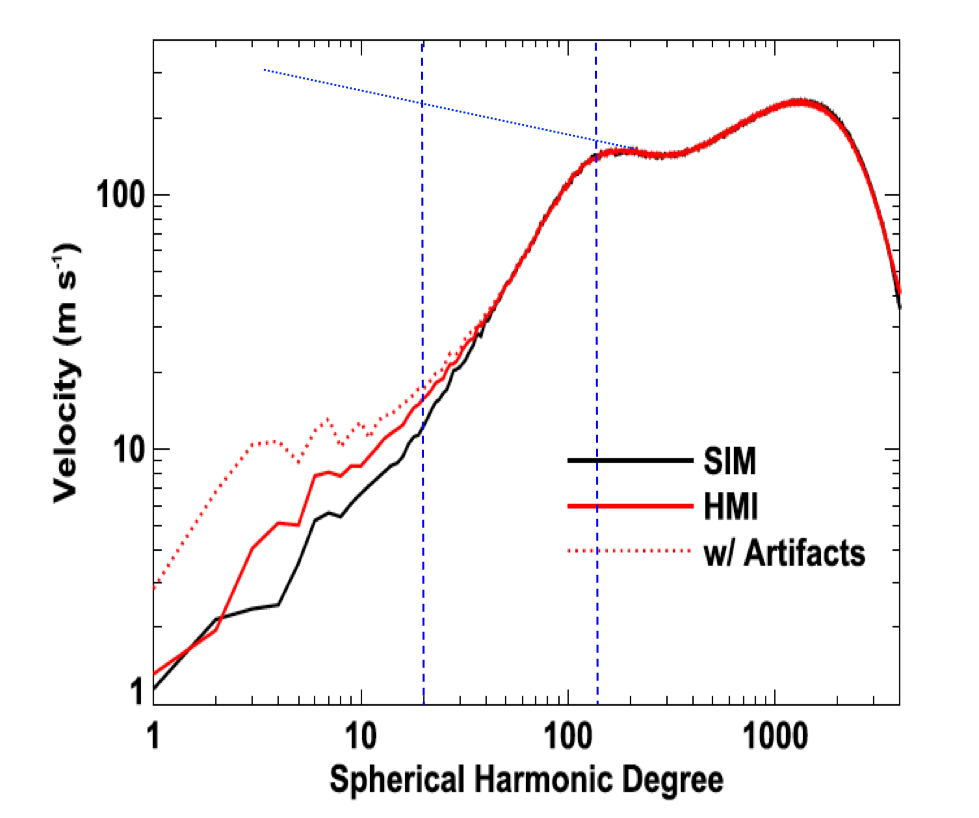}
\caption{From~\cite{2015ApJ...811..105H}: Solar Doppler velocity spectrum as determined from Helioseismic and Magnetic Imager (HMI) observations ({\it red} curves, with and without removal of image artifacts), along with a random-phase synthetic spectrum ({\it black} curve). Vertical {\it blue dashed} fiducial lines have been added indicating the approximate scale of supergranulation and giant-cells. The {\it blue dotted} line approximately over-plots the spectrum seen in numerical simulations. 
Figure without added {\it dashed} and {\it dotted blue} lines used with permission of~\cite{2015ApJ...811..105H}.}
\label{fig_rast_hathawayspectrum}
\end{figure}

These issues are critical to our understanding of large-scale motions on the Sun. As can be seen in Figure~\ref{fig_rast_hathawayspectrum} (from~\cite{2015ApJ...811..105H}), only two of the components described above are evident as distinct features in the observed spectrum of motions in the solar photosphere.  Granulation is responsible for the most pronounced peak at high spherical-harmonic degree and supergranulation for the smaller peak near spherical-harmonic degree 120.  Added to the plot are vertical fiducial lines indicating the approximate scale of supergranulation and giant cells.  Additionally, a {\it blue dotted} line has been added to schematically indicate the monotonic increase of power to low wavenumbers seen in all realistic numerical simulations up until the most recent of~\cite{hotta2021}. In the~\cite{hotta2021} simulations in the power rolls over at spherical-harmonic degree $\sim10$.  It is the discrepancy in low spatial-frequency power between simulations and observations that has come to be known as the convective conundrum.  

It is important to note that the spectrum plotted in Figure~\ref{fig_rast_hathawayspectrum} is a composite, with vertical velocities dominating at high spatial wavenumbers (granular scales) and horizontal motions most important at supergranular scales.  The vertical velocity contribution decreases from the granular peak towards lower wavenumbers, with horizontal velocity contribution increasing to spherical-harmonic degree $\sim120$ before rolling over beyond that.  The supergranular peak results from this decrease in the power beyond spherical-harmonic degree $\sim120$.  Thus, with respect to photospheric flow observations, the convective conundrum refers to the scale and amplitude of the horizontal-flows in the photosphere.  No global-spherical-shell or local-area simulation of solar convection yet captures the supergranular scale maximum in photospheric horizontal-flow power.

\subsection{Solar inertial modes}

On scales much larger than supergranule, non-axisymmetric flows in the Sun 
have long timescales and are thus strongly influence by solar rotation via the Coriolis force.
A component of these flows has recently been observed as waves of radial vorticity at the surface and are known as inertial modes.
In this section, we give an overview of their characteristics and properties.

Inertial modes owe their existence to rotation \citep[][]{greenspan1968}. 
Their restoring force is the Coriolis force. 
In a uniformly rotating sphere, the frequencies of inertial modes are limited to a range of $\lvert \omega \rvert <2\Omega_{0}$ in the co-rotating frame, where $\Omega_{0}$ denotes the angular velocity. 
The traditional Rossby modes (or r modes) correspond to a variety of inertial modes that have quasi-toroidal motions.
Although these modes have been expected to exist in the Sun and stars since the late 1970's \citep[e.g.,][]{papaliozou1978,saio1982,unno1989}, they were not observed on the Sun until very recently.
Inertial modes on the Sun have very long oscillation periods (of the order of months) and very small velocity amplitudes (of the order of 1 m~s$^{-1}$). 
Therefore, long-term and high-precision observations of horizontal flows over many years are required to detect them.

\citet[][]{loeptien2018} discovered the solar equatorial Rossby modes using both a granulation-tracking method and ring-diagram analysis applied to six years of SDO/HMI data.
These Rossby modes are retrograde-propagating waves of radial vorticity.
\citet[][]{loeptien2018} found the excess power in the radial vorticity along the dispersion relation of the sectoral Rossby modes, $\omega=-2\Omega_{\mathrm{eq}}/(m+1)$, for azimuthal orders $3 \leq m \leq 15$.
In this formula, $\Omega_{\mathrm{eq}}$ is the equatorial rotation rate of the Sun at the surface.
For $ m \gtrsim 5$, the latitudinal eigenfunctions of the equatorial Rossby modes significantly deviate from the sectoral spherical harmonics: the radial vorticity peaks at the equator but changes sign at middle latitudes (this is due to differential rotation, see below).
The detection of solar equatorial Rossby modes has been confirmed in follow-up studies using various other observational datasets and methods \citep[][]{liang2019,proxauf2020,mandal2020,hanson2020,2021ApJ...908..160H}.

The observed equatorial Rossby modes exist only at very large scales with $m \leq m_{\mathrm{crit}}=15$.
\citet{loeptien2018} speculated that this critical azimuthal order $m_{\mathrm{crit}}$ might reflect the Rhines scale $l_{\mathrm{rhines}}=\sqrt{R_{\odot} v_{c}/\Omega_{\odot}}$ above which rotation strongly affects  turbulent convection \citep[][]{rhinesWavesTurbulenceBetaplane1975}.
Assuming $m_{\mathrm{crit}} \approx R_{\odot}/l_{\mathrm{rhines}}$, a typical speed of turbulent convection can be roughly estimated as $v_{\mathrm{c}} \approx 9$ m~s$^{-1}$, which is about one order magnitude smaller than the typical  mixing-length estimate \citep[][]{1958ZA.....46..108B,stix2002}. 
This may add another piece of evidence to the solar convective conundrum  (see Sections 2.2 and 3.1 below).

\begin{figure}[t]
\centering
\includegraphics[width=0.99\textwidth]{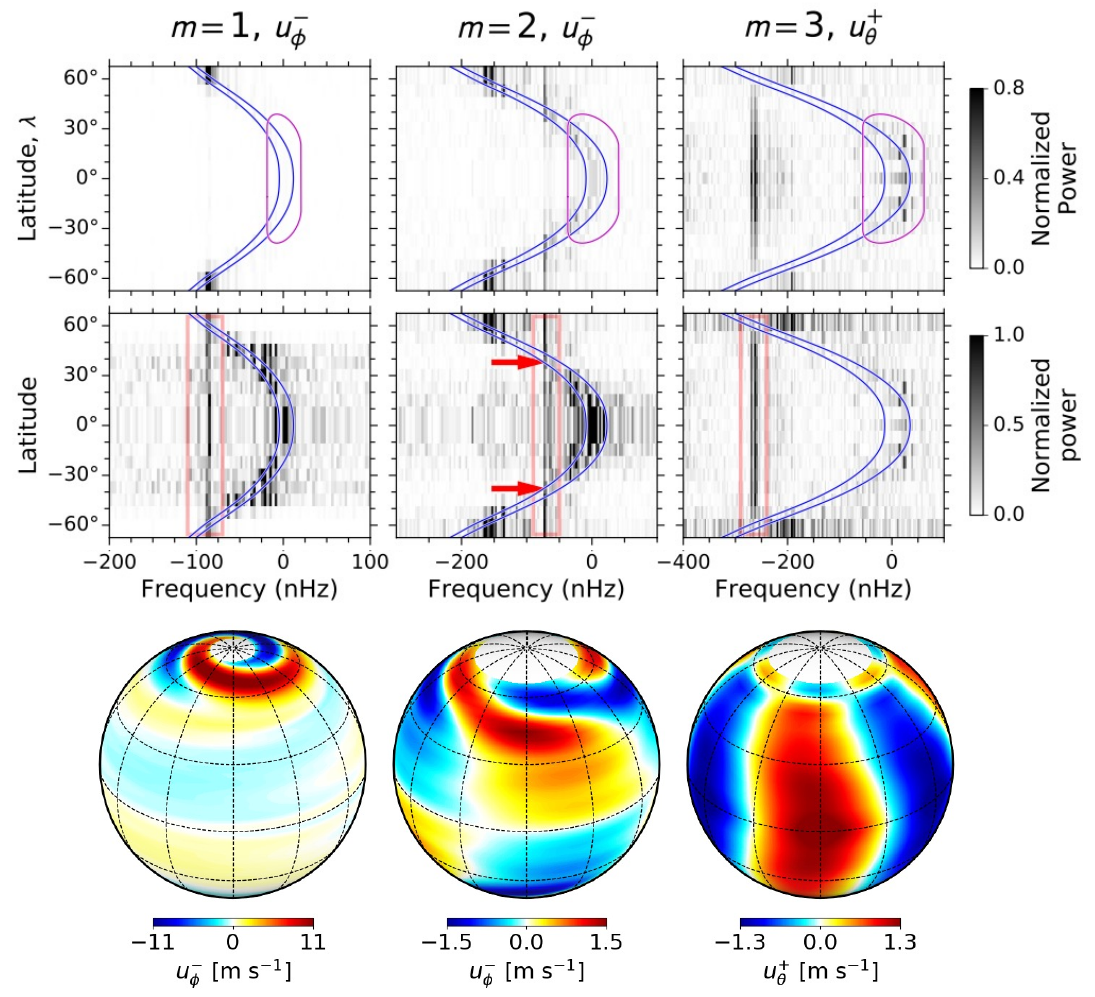}
\caption{
Observational power spectra in the Carrington frame  and eigenfunctions of three selected inertial modes of the Sun.
\textit{Top row}: Power spectra of the longitudinal component of velocity $u_\phi$ for $m=1$ (left column) and  $m=2$ (middle column), and power spectrum of the colatitudinal component of velocity $u_\theta$ for $m=3$.
The blue curves show the differential rotation rate at $r=0.96R_{\odot}$ and $R_{\odot}$.
The purple contour indicates the region  affected by active-region flows.
\textit{Middle row}: The same power spectra but normalized at each latitude by their average value over the frequency range between the orange bars. Excess power is seen at a specific frequency at all latitudes in each of the three cases. The red arrows point to critical latitudes at the surface for the case $m=2$.
\textit{Bottom row}: Observed horizontal velocity eigenfunctions at the surface for the $m=1$ high-latitude mode, the $m=2$ critical-latitude mode, and the $m=3$ equatorial Rossby mode. These figures are courtesy of \citet[][]{gizon2021}.
}\label{fig_bekki:inertial}
\end{figure}

Recently, \citet[][]{gizon2021} analyzed more than 10 years of data from both SDO/HMI and GONG and detected additional quasi-toroidal inertial modes with $1 \leq m \leq  10$.
With the help of a 2.5D linear eigenvalue solver applied to a model of the differentially rotating convection zone   \citep[][]{bekki2022a}, they identified not only equatorial Rossby modes, but also  modes at middle and high latitudes.
The observational power spectra and the measured eigenfunctions of three selected  inertial modes with $m=1, 2$, and $3$ are shown in Fig.~\ref{fig_bekki:inertial}.
These modes are very sensitive to the solar latitudinal differential rotation \citep[see][for a discussion in the $\beta$ plane]{gizon2020a}. For many of the inertial modes, there exists latitudes at which the phase speed is equal the local differential rotation speed; such latitudes are called critical latitudes (see Fig.~\ref{fig_bekki:inertial}, middle column). 
The $m=1$ high-latitude mode has a large amplitude ($v_{\phi} \approx 10-20$ m~s$^{-1}$) at high latitudes (above $\approx 50^{\circ}$) and its surface eigenfunction exhibits a spiral pattern in the polar regions. The horizontal flow associated with the $m=1$ high-latitude mode was first observed by \citet{2013Sci...342.1217H}, but misidentified at the time as giant cell convection.
This phenomenon have also been observed by \citet[][]{bogart2015} and \citet[][]{howe2015}.
A linear model suggests that this $m=1$ high-latitude mode is self-excited when a large enough latitudinal entropy gradient exists in the convection zone \citep[][]{bekki2022a}. 

More recently, \citet[][]{hanson2022} reported the detection of another class of inertial modes with north-south anti-symmetric radial vorticity across the equator.
These modes propagate in a retrograde direction with the phase speed roughly three times faster than the that of the equatorial Rossby modes.
Compared with the equatorial Rossby modes and the high-latitude modes reported in \citet[][]{gizon2021}, these so-called \textit{high-frequency retrograde} modes have lower velocity amplitudes and are thus much harder to distinguish from the background noise in the power spectra.
According to simplified linear eigenmode calculations of a uniformly-rotating Sun \citep[][]{triana2022,bhattacharya2023}, these modes are not quasi-toroidal, i.e., substantial radial motions are involved.

Though it remains unclear how important the solar inertial modes are to the overall convection zone dynamics, they play an important diagnostic role.
\citet[][]{gizon2021} and \citet[][]{bekki2022a} have shown that the properties of some inertial modes (i.e., frequencies, linewidths, and surface eigenfunctions) are sensitive to the turbulent viscous diffusivity $\nu_{\mathrm{t}}$ and to  the superadiabaticity $\delta$ of the convection zone.
\citet[][]{gizon2021} inferred that, on average,  $\nu_{\mathrm{t}} \leq 10^{12}$ cm$^{2}$~s$^{-1}$ and $\delta < 2\times 10^{-7}$. These values are an order of magnitude smaller than the theoretical estimates from the local mixing length model \citep[][]{christensen-dalsgaardCurrentStateSolar1996,munoz2011}. 
It is noteworthy that both of these parameters cannot be constrained by conventional p-mode helioseismology, and are important in discussions of the convective conundrum (Sections 2.2 and 3.1 below) and the solar dynamo. 
The amplitudes of the linearly-stable modes might provide additional constraints on the turbulent viscosity \citep{philidet2023}.
On the other hand, to better understand the amplitudes of the linearly-unstable inertial modes, nonlinear numerical simulations will be required \citep[][]{bekki2022b,matilsky2022,bekki2023}.

\subsection{Observations of large-scale flows on solar-type stars}
\label{sec:stell_large_scale_obs}

It is possible to obtain  some information about the rotation of distant stars  using observations of photometric variability, radial velocity measurements, and asteroseismology. Such methods are challenging, however recent results are setting new constraints on stellar differential rotation. 

One possibility is to measure the effects of rotation on stellar oscillations. Rotation lifts the  degeneracy of the frequencies of the modes of oscillation with different azimuthal orders. Latitudinal and radial differential rotation may leave a signature in the fine structure of the oscillation power spectra. See, e.g., \citet{Aerts2010} and  \citet{garciaAsteroseismologySolartypeStars2019} for reviews of  \textit{asteroseismology}. 
In the last years it has become possible to probe the deep interior of some evolved stars, including red giants \citep{2012Natur.481...55B,deheuvelsSEISMICEVIDENCERAPIDLY2012} and, more recently, subgiants \citep{deheuvelsSeismicEvidenceSolidbody2020}. 
Probing the slowly-rotating Sun-like stars is especially difficult because rotational splitting frequencies are often too small compared to the mode linewidths. 
Constraints on the mean angular velocity and the inclination angle of the rotation axis have been obtained in a few cases \citep[e.g.,][]{Gizon2013,Chaplin2013}.
Strong latitudinal differential rotation has recently been observed on a few selected main-sequence solar-type stars  \cite{benomarAsteroseismicDetectionLatitudinal2018}. 
Solar-like differential rotation profile (equator faster than the poles) is detected on 13 stars with a much stronger amplitude than on the Sun (on average twice the solar value). Subsequently, \cite{bazotLatitudinalDifferentialRotation2019} reported latitudinal rotation contrasts closer to the solar value for the two solar analogs 16~Cyg~A and~B. The equator-to-pole difference in angular velocity, $\Delta\Omega$, may depend on the mean rotation rate of the stars, with the majority of stars in \cite{benomarAsteroseismicDetectionLatitudinal2018}'s study rotating faster than the Sun.
Only upper limits have been set on the radial differential rotation of Sun-like stars \citep{Nielsen2014}. To understand differential rotation in stars, it is important to study its relationship to the mean rotation rate $\Omega_*$, the stellar mass $M_*$, the stellar age, and the stellar composition.

Surface rotation and differential rotation can also be studied using observations of stellar photometric variability. The presence of starspots and faculae on rotating stars modulate the photometric signal at the rotation period \citep[e.g.,][]{Nielsen2013,Reinhold2022}. Some information about latitudinal differential rotation can be retrieved when magnetic active regions are present over a range of latitudes on the stellar surface \citep[e.g.,][]{Reinhold2013,Nielsen2019}.
The stellar differential rotational versus mean rotation rate may be described by a  power law, $\Delta\Omega\propto\Omega_*^n$. Figure~3 of \cite{barnesDependenceDifferentialRotation2005} summarizes the different studies that were available at the time. The long-term monitoring of photometric modulations point toward a weak rotational dependence, $n=0.24$ \citep{1995ApJS...97..513H}, while similar observations in H \& K bands of Ca II emission suggest a significantly higher value, $n=0.7$ \citep{1996ApJ...466..384D}.  The spectroscopic analysis of \citet{reinersRotationDifferentialRotation2003} gave a  power law index of $n=0.66$.
\cite{barnesDependenceDifferentialRotation2005} proposes an index of $n=0.15$, but also show that this index is very sensitive to the spectral-type diversity of the target sample considered~\citep[see also][]{reinersRotationDifferentialRotation2003}.  For instance, \cite{reinholdRotationDifferentialRotation2013}  find a value of $n=0.3$ for cool stars and  \cite{balonaDifferentialRotationStars2016}  report $n=0.2$ for G-stars. Asteroseismic studies tend to find values for $n$ between $0.3$ and $0.45$ \citep{garciaAsteroseismologySolartypeStars2019},  and underline its sensitivity to the effective temperature $T_{\rm eff}$ range of the stars considered  \citep[see][and references therein]{reinholdRotationDifferentialRotation2015}.

The dependence of the latitudinal contrast $\Delta\Omega$ on spectral type also appears to be significant for fast main-sequence rotators.  \cite{colliercameronDifferentialRotationRapidly2007} studies fast M, K, G, and F main-sequence rotators and reports a strong $\Delta\Omega\propto T_{\rm eff}^{8.6}$ dependence on the effective temperature, consistent with \cite{barnesDependenceDifferentialRotation2005}. Large error bars are associated with the hottest spectral types, however the models of \cite{kuekerDifferentialRotationMeridional2011} tend to confirm a strong dependence \citep[see][for a detailed discussion]{reinholdRotationDifferentialRotation2015}.

To take into account both rotational and spectral-type aspects, \cite{saarStarspotsCyclesMagnetic2010} proposed to consider the global shear $\Delta\Omega$ as a function of the stellar Rossby number $Ro_{\rm s}=\tau_{\rm c}/\Omega_*$ (see also \citealt{brunDifferentialRotationOvershooting2017} and \citealt{norazHuntingAntisolarDifferentially2022} for definitions and prescriptions). Indeed, it is possible to parameterize the convective turnover time $\tau_{\rm c}$ as a function of $T_{\rm eff}$ \citep{cranmerTESTINGPREDICTIVETHEORETICAL2011}. In particular, he finds that $\Delta\Omega\propto Ro_{\rm s}^{-1}$ for unsaturated rotators ($\Omega\leq 12\Omega_\odot$), pointing then toward $n=1$ when fixing $T_{\rm eff}$ and the composition. Numerical studies of the last decade have confirmed that the Rossby number is a major parameter to consider regarding the characterization of large-scale flows along stellar evolution (see Section~\ref{sec:stellar_modeling}). However, it has to be mentioned here that the high-Rossby regime still needs observations to constrain theoretical results. A difficulty lies here in the sensitivity of the aforementioned techniques to the rotation rate, which decreases signals significantly when considering slow-rotators.

Finally, recent observational studies have started to investigate the impact of the metallicity, for instance with the solar analog HD 173701 studied by \citet{karoffInfluenceMetallicityStellar2018}. Its parameters are indeed close to the solar ones, while having a significantly higher metallicity ([Fe/H]$=0.3\pm 0.1$). Using different methods previously mentioned, the authors report a solar-like differential rotation (fast equator) with a latitudinal contrast being twice the solar one. Monitoring of its chromospheric and photometric emissions also show a cyclic activity shorter than the one observed on the Sun ($P_{\rm cyc}=7.4$ against 11 years), while having a higher amplitude of variation, which underlines the entanglement between composition, large-scale flows, and magnetism \citep{brunMagnetismDynamoAction2017,seePhotometricVariabilityProxy2021}.

To summarize, differential rotation is a characteristic quantity of stellar convective envelopes. Apart from the Sun, the exact quantification of the surface rotational contrast remains difficult because it requires a high degree of precision of the instruments used and long acquisition periods for the different targets. New observations by the upcoming PLATO mission should allow major advances in this direction \citep{rauerPLATOMission2014}. A particular focus for future prospects will lie on the monitoring and characterization of slow-rotators, which appear to be challenging targets for current techniques (\citealt{norazHuntingAntisolarDifferentially2022,2023arXiv230714190D}, see also Section~\ref{sec:stellar_modeling}). In the meantime, theoretical modeling of these flows are hence crucial for the understanding and support of these observations, and in order to guide those to come.

\section{Models of large-scale flows} \label{sec:formation_of_large-scale_flows}

Large-scale flows in the Sun are generated and maintained by thermal convection.  In this section we briefly summarize the current theoretical understanding of how that occurs.

\subsection{Mixing length theory and energy budget}

Mixing-length theory~\citep[MLT;][]{1958ZA.....46..108B} remains a remarkably useful way to describe the mean energy transport by convection even when the actual dynamics are far from localize eddy motions. 
In the mixing length formulation, we can relate energy flux to the convective velocity and the stratification through a parameter called the mixing-length parameter $\alpha_\mathrm{MLT}=L_\mathrm{MLT}/H_p$, where $L_\mathrm{MLT}$ and $H_p=-\left(d\log p/dr\right)^{-1}$ are the mixing length and the pressure scale height in the convecting fluid. The mixing-length parameter, $\alpha_\mathrm{MLT}$ is a parameter of order unity which mainly affects the stellar radius \citep{demarque_1964ApJ...140..541D} when used in a stellar structure model. Here we take $\alpha_\mathrm{MLT}=1$ (i.e., $L_\mathrm{MLT}=H_p$) for the simplest mixing-length formulation.  
The typical temperature perturbation in the convection $\Delta T$ is then evaluated as:
\begin{align}
  \Delta T &= \left[\left(\frac{dT}{dr}\right)_\mathrm{ad} - \frac{dT}{dr}\right]L_\mathrm{MLT} \nonumber 
  = T\left[\left(\frac{d\log T}{d\log p}\right)_\mathrm{ad} - \left(\frac{d\log T}{d\log p}\right)\right] \nonumber \\
  &   =  - T \delta\ , 
\end{align}
where $\delta = (d\log T/d\log p) - (d\log T/d\log p)_\mathrm{ad}$ is the superadiabaticity, the deviation from the adiabatic stratification. When the speed of sound in the medium is much faster than convection, the sound wave instantaneously relaxes any pressure perturbation $\Delta p$, and the density perturbation $\Delta \rho$ can be 
estimated from the linearized equation of state as
\begin{align}
  \frac{\Delta \rho}{\rho} = - \frac{\Delta T}{T} = \delta\ .
  \label{eqn_mlt1}
\end{align}
A simplified equation of motion is converted with rough dimensional analysis as
\begin{align}
  \rho \frac{D v}{Dt} = - \Delta \rho g \rightarrow \frac{v_\mathrm{c}}{\tau_\mathrm{c}} = -\frac{\Delta \rho}{\rho}g,
\end{align}
where $v_\mathrm{c}$ is the typical convection velocity. Here we can evaluate the typical time for the convection as $\tau_\mathrm{c}=H_p/v_\mathrm{c}$. Then the convection velocity can be written as 
\begin{align}
  v_\mathrm{c}^2 = \delta \frac{p}{\rho} \sim \delta c^2_\mathrm{s}\ ,
\end{align}
where $c_\mathrm{s}$ is the speed of sound. 
Together, these approximations yield several important approximations for the superadiabaticity: 
\begin{align}
  \delta \sim \left\vert \frac{\Delta T}{T} \right\vert \sim  \left\vert \frac{\Delta \rho}{\rho} \right\vert \sim \left(\frac{v_\mathrm{c}}{c_\mathrm{s}}\right)^2,
  \label{eqn_mlt}
\end{align}
While some more careful treatments include additional physical effects, such as variation in the mean molecular weight of the plasma or the fluid drag force, in the practical application of MLT, the relationships captured by Equation ~\ref{eqn_mlt} change little.  
What is important is that these relations lead to a ratio of the kinetic energy $E_\mathrm{kin}$ to the internal energy $E_\mathrm{int}$ that scales with superadiabaticity $\delta$:
\begin{align}
  \frac{E_\mathrm{kin}}{E_\mathrm{int}} \sim \frac{\rho v_\mathrm{c}^2/2}{\rho c_\mathrm{v}T} \sim \left(\frac{v_\mathrm{c}}{c_\mathrm{s}}\right)^2 \sim \delta\ .
\end{align} \par

In the convection zone, heat is mainly transported by the convection, with the enthalpy flux given by 
\begin{align}
  F_e = \rho c_\mathrm{p} \Delta T v_\mathrm{c} \sim \rho v_\mathrm{c}^3 \ .
\end{align}
Normalizing this equation yields
\begin{align}
  \frac{F_e}{\rho c_\mathrm{s}^3} \sim \left(\frac{v_\mathrm{c}}{c_\mathrm{s}}\right)^3 \sim \delta^{3/2}\ .
  \label{eqn_mlt2}
\end{align}
Over the depth of the convection zone, with the superadiabaticity taken to $10^{-6}$ and $10^{-1}$ at the base and surface respectively \citep[e.g.,][]{Ossendrijver2003}, the convective velocity amplitudes should change by a factor of a few hundred. At the base of the solar convection zone, the flow is subsonic (Mach number $\sim 10^{-3}$) and the internal energy of the fluid is much larger ($10^6$ times larger) than the kinetic energy of the flows.

\subsection{The effect of stratification on observed horizontal convective flow amplitudes}

As is clear from the mixing-length calculation above, one of the most important aspect of stellar envelope convection is the steep stratification of the mean state.  The solar convection zone is about $210~$Mm deep, over which the density changes by a factor of about one million and the pressure by about 800 million.  The density scale height in the photosphere is about $150~$km, while at the convection zone base it is equal to nearly half the depth.  This has profound influence on the convective dynamics.

By mass conservation, only a very small fraction of the upwelling fluid from the deep convection zone makes it into the photosphere.  The rest must overturn.  Over each scale height, the density decreases by a factor of $1/e$, so that $1-1/e$ of the mass must overturn.  Similarly, the downwelling fluid must entrain mass at this rate.  For simple assumptions about the flow geometry, this implies a characteristic horizontal flow scale at each depth $d=4H_\rho$, where $H_\rho$ is the density scale height~\citep{ 2009LRSP....6....2N}.  Taking this to be the integral (driving) scale of the motions~\citep{2009AIPC.1094..764S} allows a simple two component model that can reproduce the observed spectrum of horizontal motions in the solar photosphere~\citep{lord2014}.

For statistically steady motions and small horizontal density gradients compared to the mean vertical stratification, the equation of mass continuity takes the form 
\begin{align}
\nabla_\mathrm{h}\cdot{\bm v}_\mathrm{h}=-{\frac{\partial v_z}{\partial z}}-{\frac{v_z}{H_\rho}}\ .
\label{eqn_anelastic}
\end{align}
This suggest two flow regimes.  For $\partial v_z/\partial z \gg v_z/H_\rho$ the motions are nearly divergenceless and for $\partial v_z/\partial z \ll v_z/H_\rho$ the vertical stratification dominates.  In these two limits,  Equation~\ref{eqn_anelastic} can be used to determine the horizontal velocity power spectrum given that of the vertical velocity.  For small scale motions, high horizontal wavenumbers $k_\mathrm{h}$, the motions are nearly isotropic with
\begin{align}
\widetilde{{\bm v}}_\mathrm{h}^* \cdot\widetilde{\bm v}_\mathrm{h}=\widetilde{v}_z^*\widetilde{v}_z\ ,
\label{eqn_highwavenumber}
\end{align}
where $\widetilde{{\bm v}}_\mathrm{h}^* \cdot\widetilde{\bm v}_\mathrm{h}$ is the power spectrum of the horizontal flows and $\widetilde{v}_z^*\widetilde{v}_z$ is that of the vertical flows.  
For low wavenumber components of the flow, on the other hand, stratification is important and
\begin{align}
\widetilde{{\bm v}}_\mathrm{h}^* \cdot\widetilde{\bm v}_\mathrm{h} ={\frac{2}{k_\mathrm{h}^2H_\rho^2}}\widetilde{v}_z^*\widetilde{u}_z\ .
\label{eqn_lowwavenumber}
\end{align}
The cross over between these two regimes occurs at the integral scale $4H_\rho$ at each depth.  Figure~1 of~\cite{lord2014} confirms that this two-component continuity balance determines the relationship between the vertical and horizontal-velocity power spectra in radiative hydrodynamic simulations.

Using this balance, a model of the horizontal motions observed in the solar photosphere can be constructed from the vertical velocity spectrum at each depth. For example, the high-wavenumber vertical-velocity spectrum can be taken to be Kolmogorov above the integral scale $4H_\rho$, with a mixing-length approximation to determine the total integrated power.  Since no scales larger than that are driven at any given depth, the amplitudes of modes with scales larger than the driving scale at any depth can be determined by their decay with height from the depth at which they were last driven. A  potential flow approximation on these scales can be used to model that amplitude decrease with height.  With these ingredients, working from the bottom of the convection zone upwards, the power spectrum of the horizontal velocity at each depth can be determined. It matches that seen in three-dimensional radiative hydrodynamic simulations~\citep{lord2014}. 
The broader and critical take away from this simplified approach is that the horizontal-velocity spectrum in the photosphere depends on the vertical-velocity flow amplitudes at depth.  The larger the horizontal flow scale observed in the photosphere, the deeper it originates.  This suggests that the dramatic decrease in the observed horizontal-velocity  power above the supergranular scale on the Sun reflects weak convective driving at depth (at depths below $\sim10$~Mm) and that supergranulation represents the largest buoyantly driven scale of motion~\citep{lord2014, 2016ApJ...829L..17C}.  

A number of reasons for the low-convective amplitudes at depth are possible, including highly non-local dynamics (maintenance of small scale downflowing plumes generated in the photosphere with little horizontal diffusion) that ensures that the mean stratification of the solar convection zone is closer to adiabatic than numerical models can achieve~\citep{2016ApJ...829L..17C, 2020ASSP...57..149R}, this possibly due to the presence of small scale magnetic field~\citep[][]{2016AdSpR..58.1475O, 2017ApJ...851...74B}; subadiabatic stratification in the deep solar convection zone due to internal heating by radiation; reduced convective amplitudes due to the stabilizing influence of rotation in the lower convection zone~\citep{2016ApJ...830L..15F, 2021PNAS..11822518V}; or a combination of these.  Any mechanism that leads to reduced convective amplitudes in the deep layers of the solar convection zone will be reflected in reduced low wavenumber power in the horizontal-flows at the surface.  

In radiative hydrodynamic simulations, a factor of $\sim2.5$ reduction in convective amplitudes below $\sim10$~Mm depth is sufficient to resolve the convective conundrum~\citep{lord2014}.  Though such a reduction has not been yet achieved in a first principles model of convection, we note that, by simple mixing-length scaling arguments (Equation~\ref{eqn_mlt}), it is consistent with the reduction in superadiabaticity ($\delta < 2\times 10^{-7}$) suggested by the~\cite{gizon2021} analysis of solar inertial modes.

\subsection{Gyroscopic pumping and thermal wind balance}

Given the convective motions, it is important to understand how, in combination with solar rotation, they generate and maintain large-scale solar differential rotation and meridional circulation.  
For that purpose, the concepts of gyroscopic pumping and the thermal wind balance are useful \citep{1999PThPh.101..189M,miesch2008,miesch_2011ApJ...743...79M}.  We discuss those in this section. For simplicity, the magnetic field is ignored, though it may be critical in some cases \citep{hotta2022}. We use a notation where a quantity $Q$ is divided into a mean (longitudinal average) $\langle Q\rangle$ and perturbation $Q'$, i.e., $Q=\langle Q\rangle + Q'$.

Gyroscopic pumping is a reflection of angular momentum conservation. The longitudinally averaged longitudinal equation of motion under the anelastic approximation can be written as
\begin{align}
  \rho_0\frac{\partial \langle\mathcal{L}\rangle}{\partial t} = - \nabla\cdot\left(\rho_0 \langle\bm{v}_\mathrm{m}\mathcal{L}\rangle\right),
  \label{eqn_angmom}
\end{align}
where  $\mathcal{L}=\varpi v_\phi$ and $\bm{v}_\mathrm{m}=v_r \bm{e}_r + v_\theta \bm{e}_\theta$ are the specific angular momentum and the meridional flow velocity, respectively, with $\varpi = r \sin \theta$ in the spherical geometry $(r,\theta,\phi)$. We also note $\rho_0$ the density background profile (spherical average). 
The flow velocity is then divided into components, as above, $\bm{v}=\langle \bm{v}\rangle + \bm{v}'$ and, assume a steady state $\partial/\partial t = 0$ balance, the equation for gyroscopic pumping can be written
\begin{align}
  \rho_0 \langle \bm{v}_\mathrm{m}\rangle \cdot\nabla \langle\mathcal{L}\rangle = -\nabla\cdot\left(\rho \varpi\langle\bm{v}'_\mathrm{m} v'_\phi\rangle\right).
  \label{equ_gyro}
\end{align}
While the angular momentum conservation equation (\ref{eqn_angmom}) determines the temporal evolution of the differential rotation, the gyroscopic pumping balance (Equation~\ref{equ_gyro}) mainly determines the meridional flow $\langle\bm{v}_\mathrm{m}\rangle$ in a steady state. The distribution of the specific angular momentum $\langle \mathcal{L}\rangle$ in the Sun is known to be mostly cylindrical, i.e., $\partial \langle\mathcal{L}\rangle/\partial z\sim0$, where $z$ denotes the direction of the rotational axis.
Thus, gyroscopic pumping can be rewritten as
\begin{align}
  \rho \langle v_\varpi\rangle \frac{\partial \mathcal{L}}{\partial \varpi} \approx - \nabla\cdot\left(\rho_0\varpi\langle \bm{v}'_\mathrm{m} v'_\phi\rangle\right).
\end{align}
Since we know that the sign of $\partial \langle\mathcal{L}\rangle/\partial \varpi$ is greater than zero, the sign of the axial torque $\mathcal{T}=- \nabla\cdot\left(\rho_0\varpi\langle \bm{v}'_\mathrm{m} v'_\phi\rangle\right)$ directly determines the direction of the meridional flow.

The thermal wind balance equation is derived from the longitudinal vorticity equation,
\begin{align}
  \frac{\partial \langle \zeta_\phi\rangle}{\partial t} =
  \langle \nabla\times\left(\bm{v}\times\bm{\zeta}\right)\rangle_\phi
  + \varpi \frac{\partial \langle\Omega\rangle^2}{\partial z} - \frac{g}{rc_p}\frac{\partial \langle s\rangle}{\partial \theta},
\end{align}
where $\bm{\zeta}=\nabla\times\bm{v}$ is the vorticity. This equation is for the evolution of the meridional flow ($v_r$ and $v_\theta$) in terms of the longitudinal vorticity $\zeta_\phi$. In a steady state ($\partial/\partial t=0$), 
\begin{align}
  \varpi \frac{\partial \langle\Omega\rangle^2}{\partial z} =
  -\langle \nabla\times\left(\bm{v}\times\bm{\zeta}\right)\rangle_\phi
  + \frac{g}{rc_p}\frac{\partial \langle s\rangle}{\partial \theta},
  \label{eqn_meanfield}
\end{align}
which reduces to the Taylor-Proudman theorem $\partial \langle\Omega\rangle/\partial z = 0$ when advection $\nabla\times(\bm{v}\times\bm{\zeta})$ and the latitudinal entropy gradient $\partial\langle s\rangle/\partial \theta$ are ignored.  Under the Taylor-Proudman constraint, contour lines of the angular velocity must be parallel to the rotational axis. As shown in Fig. \ref{fig_bekki:dr} the solar differential rotation does not follow this configuration, showing instead the prominent tachocline and the near surface shear layer at its boundaries and a more conical profile in the interior. 
The latitudinal entropy gradient plays a dominant role in forcing the rotation profile away from the Taylor-Proudman state in the deep convection zone where the rotational influence is important. 
The Reynolds stresses arising from vigorous surface convection (and the magnetic field which we ignored here) are also crucial in the near surface layer \citep[e.g.,][]{hotta2022} (see Section \ref{sec:model_large_scale_flow} for more details.)

\subsection{Governing equations for numerical simulations}

Simulating the global scale motions seen on the Sun directly requires simulating the convective motions in a spherical domain over many scale heights.  This in turn requires running efficient numerical code on high-performance supercomputers. For tractability, a number of approximations must be made during formulation.

The solar convection zone is fully ionized below about $20$~Mm, and, for global spherical shell magnetohydrodynamic models of convection below that depth, we can reliably assume the equation of state is close to that of a perfect gas. Simulations of the deep solar convection zone must include rotation along with gravitational stratification, and since the superadiabaticity in the solar convection zone is tiny ($\delta \lesssim 10^{-6}$), solving an entropy equation is preferable to formulations in terms of the total energy or internal energy. 
The set of equations to be solved can thus be written as
\begin{align*}
    \frac{\partial \rho}{\partial t} &= -\nabla\cdot\left(\rho \bm{v}\right), \\
    \frac{\partial}{\partial t}\left(\rho\bm{v}\right) &= -\nabla\cdot\left(\rho\bm{vv}\right) -\nabla p + \rho \bm{g} + \frac{1}{4\pi}\left(\nabla\times\bm{B}\right)\times\bm{B} + 2\rho\left(\bm{v}\times\bm{\Omega}_0\right), \\
    \rho T \frac{\partial s}{\partial t} &= - \rho T \left(\bm{v}\cdot\nabla\right)s + Q_\mathrm{rad}, \\
    \frac{\partial\bm{B}}{\partial t} &= \nabla\times\left(\bm{v}\times\bm{B}\right), \\
    s &= c_\mathrm{v}\log\left(\frac{p}{\rho^\gamma}\right). \\
\end{align*}
Here, we ignore the viscosity, the magnetic diffusivity, and the thermal conductivity due to the large fluid/magnetic Reynolds and Peclet numbers in the Sun.
Moreover, since perturbations in the thermodynamic variables scale with the superadiabicity (Equation~\ref{eqn_mlt1}) 
a linearized equation of state is appropriate,
\begin{align}
    \frac{s_1}{c_\mathrm{v}} = \frac{p_1}{p_0} - \gamma\frac{\rho_1}{\rho_0}\ ,
\end{align}
where $\gamma$ is the specific heat ratio and subscripts 0 and 1 denote the background and perturbed variables, such that $Q=Q_0+Q_1$, and we typically assume the hydrostatic balance for the background stratification $\rho_0$, $p_0$, and $T_0$:
\begin{align*}
    \frac{dp_0}{dr}(r) = -\rho_0 (r)g(r)\ .
\end{align*}
Pressure gradient and the gravitational forces in the momentum equation are then due to perturbations about this background state,
\begin{align}
    -\nabla p + \rho \bm{g} \rightarrow -\nabla p_1 + \rho_1 \bm{g}\ .
\end{align}

Due to a large optical depth in the deep convection zone, the diffusion approximation can be used for radiation energy transfer $Q_\mathrm{rad}$,
\begin{align}
    Q_\mathrm{rad} = - \kappa_\mathrm{rad} \nabla T,
\end{align}
where $\kappa_\mathrm{rad}$ is the radiation diffusion coefficient estimated from the local opacity.

Employing an anelastic approximation is important for deep solar convection simulations since the speed of sound $c_\mathrm{s}$ is much faster than the convection velocity $v_\mathrm{c}$. 
Explicitly solving for sound waves in the domain severely restricts the CFL condition on the time stepping $\Delta t$, and a huge number of the time steps would be required to evolve the convective motions and larger-scale flows over dynamical time scales.  
To avoid this difficulty, the anelastic approximation  \citep{1969JAtS...26..448G} is widely used \citep{cluneComputationalAspectsCode1999}, which simplifies the equation of continuity,
\begin{align}
    \nabla\cdot\left(\rho_0\bm{v}\right) = 0 \ ,
    \label{eqn_anelast}
\end{align}
filtering out sound waves by taking the sound speed to be infinite. In the context of MHD, the anelastic approximation eliminates the fast magneto-sonic waves, while preserving the Alfvén waves and the slow magneto-sonic waves.

Other sonic-filter formulations have been developed including the \textit{Lantz-Braginsky-Roberts} (LBR) method, in which a reduce pressure is introduced and interactions between fluctuating pressure and stratification are neglected ( \citealt{1992PhDT........78L}, \citealt{braginskyEquationsGoverningConvection1995}). The LBR method has the advantage of conserving energy well in both unstable convective zones and stable radiative interiors, critical for simulating gravity wave excitation and propagation (\citealt{brownENERGYCONSERVATIONGRAVITY2012},\citealt{vasilENERGYCONSERVATIONGRAVITY2013}).

Recently, a method that has come to be known as the Reduced Speed of Sound Technique (RSST) has also found extensive use.  With this, the continuity equation is altered,
\begin{align}
    \frac{\partial \rho_1}{\partial t} = - \frac{1}{\xi^2}\nabla\cdot\left(\rho \bm{v}\right)\ ,
\end{align}
so that the effective speed of sound is reduced by a factor of $\xi$ \citep{rempel_2005ApJ...622.1320R,hotta_2012A&A...539A..30H}. An advantage of the RSST method is that it does not require global communications in a parallel computing environment, in contrast to the anelastic approximation which requires frequent global communication to solve the elliptic equation~(\ref{eqn_anelast}). Thus, the RSST is very useful when solving the MHD equations with massively parallel supercomputers. Additionally, an inhomogeneous $\xi$ can be employed, taking $\xi$ large in the deep layers where the sound speed is fast and $\xi=1$ in the near-surface layer where the anelastic approximation is not valid.  This enables simulation over a continuous domain that covers the whole convection zone \citep{hotta2019}.

\subsection{Modeling of the solar large-scale flows}
\label{sec:model_large_scale_flow}

The last fifty years have brought tremendous advances in the simulation of solar and stellar convection and our understanding of the global flows that result in rotating domains.  
Initial analytic analysis of convection in a rotating sphere \citep{1961hhs..book.....C,1968RSPTA.263...93R,busseThermalInstabilitiesRapidly1970} were extended to
to numerical linear and nonlinear numerical studies aimed at understanding the behavior of rotating turbulent astrophysical bodies~\citep{1975JAtS...32.1331G,1977GApFD...8...93G}. 
In particular, the aim was to understand the importance of 
nonlinear process in both the solar~\citep{gilmanModelCalculationsConcerning1979} and terrestrial~\citep{cuongGenerationMagneticFields1981} contexts.  In the solar case, this was strongly 
motivated by the need to explain solar differential rotation, which had at the time been observed for more than one century \citep{1860MNRAS..20..254C}. 

The earliest numerical calculations were done using the Boussinesq approximation, but quickly extended to include stratification using the anelastic approximation \citep{gilmanCOMPRESSIBLECONVECTIONROTATING1981,1982ApJ...256..316G}. At the same time, dynamo calculations were solving for the evolution of a magnetic field in these simulations, adding new theoretical constraints on the understanding of its generation within the Sun \citep{gilmanDynamicallyConsistentNonlinear1981,glatzmaierNumericalSimulationsStellar1984,1985ApJ...291..300G}.  In parallel,
stellar evolution models became sophisticated enough to model in detail the solar stratification (i.e., density, pressure, temperature, equation-of-state, opacity as functions of depth) and these models were found to be highly consistent with deductions based on helioseismic measurements~\citep[Model S:][]{christensen-dalsgaardCurrentStateSolar1996}.
With the stratification well modeled, notable advances could be made in a more realistic reproduction of thermal convection \citep[e.g., ][]{miesch2000,brunTurbulentConvectionInfluence2002}. Although spatial resolutions were moderate ($N_\theta < 128$, $\ell_{\rm max}<85$), the solar-like differential rotation profile (with faster equator and slower poles) was reproduced in these anelastic studies.\par

However, the differential-rotation profiles found in numerical solutions tended to obey the Taylor-Proudman theorem, i.e., $\partial \langle\Omega\rangle/\partial z=0$, in contrast to the solar observation (Fig.~\ref{fig_bekki:dr}). Motivated by a thorough assessment of mean-field dynamics~\citep{rempel_2005ApJ...622.1320R}, \cite{mieschSolarDifferentialRotation2006} adopted latitudinal entropy gradient at the bottom boundary and achieved non-Taylor-Proudman differential rotation, as indeed is suggested by Equation~\ref{eqn_meanfield}. This somewhat ad-hoc method was adopted by several follow-up studies \citep[e.g.][]{miesch2008,fan_2014ApJ...789...35F}. In particular, \citet{brun_2011ApJ...742...79B} included the overshoot layer between the convection zone and the deeper radiative layer below in a self-consistent model, suggesting that the interaction between the two layers is responsible for the crucial latitudinal entropy gradient, the tachocline, and the conical profile observed in the bulk of the convection zone. 

This conclusion has not gone without debate.  Earlier, \cite{miesch2000} concluded that anisotropic entropy transport by the overshooting downflows may not be enough to maintain a solar-like differential rotation profile, though \cite{hotta_2018ApJ...860L..24H} point out that an efficient small-scale dynamo can amplify the effect \citep[see also][]{hotta_2015ApJ...803...42H}, and help produce required non-Taylor-Proudman state. Recently \cite{matilsky2020} have pointed out that the latitudinal entropy gradient achieved is sensitive to the radial boundary condition imposed. Resolution of these uncertainties is critical to our understanding of the origin of the observed radial profile of the solar differential rotation.\par

The other non-Taylor-Proudman feature in the solar convection zone is the near-surface shear layer (NSSL). 
The NSSL is thought to be generated by the radially-inward angular momentum transport by near-surface convection which is not strongly influenced by rotation \citep[e.g.,][]{foukal1975}.
Numerical simulations have succeeded in reproducing part of the observed feature \citep{guerreroDIFFERENTIALROTATIONSOLARLIKE2013,hotta2015,matilsky_2019ApJ...871..217M}, with turbulent viscosity playing an important role \citep{hotta2015}, however, the models fail to reproduce key aspects of the NSSL, especially at  mid-latitude. \cite{matilsky_2019ApJ...871..217M} argue that the role of large-scale columnar convection (banana cells) and meridional circulation and their balance differ with latitude, but are unable to produce a solution that captures the NSSL in both the equatorial and high-latitude regions. Overall, a consensus has not been reached on the maintenance mechanism for the NSSL.\par

\subsection{Towards modeling other stars}
\label{sec:stellar_modeling}

Using models such as the ones presented previously, numerical simulations are a powerful tool to study the formation and dynamics of large-scale flows in the astrophysical context. In particular, full sphere simulations, resolving a broad range of turbulent convective scales, are suitable tools to probe the large-scale dynamics of distant stars. As examples, numerical studies of solar-type stars recently provided new constraints on the trends we can expect for differential rotation ($\Delta\Omega\propto\Omega_*^{0.46}$) on G and K stars \citep{brunPoweringStellarMagnetism2022}, and  highlight the different large-scale flows regimes possible. These regimes can be characterized by the Rossby number \citep{gastineSolarlikeAntisolarDifferential2014,brunMagnetismDynamoAction2017,hindmanMorphologicalClassificationConvective2020}.

Thanks to numerical simulations \citep{mattConvectionDifferentialRotation2011,guerreroDIFFERENTIALROTATIONSOLARLIKE2013,kapylaConfirmationBistableStellar2014,simitevDYNAMOEFFECTSTRANSITION2015,brunDifferentialRotationOvershooting2017,karakConsequencesHighEffective2018}, three regimes are currently acknowledged in global rotating models of main-sequence solar-type stars:
1. At low Rossby numbers, differential rotation profiles are highly constrained (Taylor-Proudman) with flows that become cylindrical \citep{1975JAtS...32.1331G,2000ApJ...533..546E,mieschSolarDifferentialRotation2006,brownRapidlyRotatingSuns2008}. In extreme cases, such profiles show alternating prograde and retrograde zonal jets, also called Jupiter-like jets \citep{rhinesWavesTurbulenceBetaplane1975,heimpelSimulationDeepseatedZonal2016}. The meridional circulations under these conditions is typically multicellular in each hemisphere and aligned along the vertical axis (see for example \citealt{brunDifferentialRotationOvershooting2017}). When magnetic fields are included in the calculation, the Lorentz force feedback can quench the flows, \citep{brunInteractionDifferentialRotation2004,yadavEXPLAININGCOEXISTENCELARGESCALE2015}, resulting in a significant decrease of the differential rotation contrast ($\Omega$-quenching) and the appearance of trans-equatorial meridional-circulation cells \citep{brunPoweringStellarMagnetism2022}.
2. At intermediate Rossby numbers, the differential rotation adopts a typical solar-like conical profile, with a fast equator and slow poles (see for instance \citealt{mieschSolarDifferentialRotation2006,hotta2022}). In this regime the power sustaining the differential rotation contrast can reach tens of percent of the stellar luminosity \citep{brunPoweringStellarMagnetism2022}.
3. At high Rossby number (typically over the unity), differential-rotation profiles in simulations become “anti-solar” \citep{gastineSolarlikeAntisolarDifferential2014}, showing then a slow equator and fast poles \citep{1977GApFD...8...93G}.

As stars spin down along the main sequence \citep{skumanich72,galletImprovedAngularMomentum2013,ahuirMagneticTidalMigration2021}, its rotational influence changes and so does by definition its Rossby number which will increase. The resulting transitions between these large-scale flow regimes may then influence the evolution of the star \citep{metcalfeOriginWeakenedMagnetic2022}, and in particular play important roles in global dynamo processes \citep{strugarekSensitivityMagneticCycles2018,warneckeDynamoCyclesGlobal2018,guerreroWhatSetsMagnetic2019}.
Large-scale flows indeed strongly influence the nature of such dynamo processes, however the Lorentz force feedback on these flows are likely to play an important role, both in the construction of the solar-type differential rotation profile \citep{hotta2022} and magnetic cycle emergence \citep{gilmanDynamicallyConsistentNonlinear1983,augustsonGRANDMINIMAEQUATORWARD2015}. Recent studies highlighted that the latter behavior may arise in specific stellar parameter ranges \citep{strugarekSensitivityMagneticCycles2018}, where the back-reaction of the Maxwell stress modulates the latitudinal differential rotation contrast, and thus the large-scale magnetic field generation. Probing of these modulations (torsional oscillations) have recently been constrained on the Sun (see Section~\ref{sec:solDRMC}) and quantified for other stars in numerical parametric studies \citep{brunPoweringStellarMagnetism2022}.

Recent observations show that the composition of the star too plays a role in the convective dynamo  \citep{seePhotometricVariabilityProxy2021}. One-dimensional stellar models show varying sensitivities to metallicity (\cite{amardImpactMetallicityEvolution2020}, see also Section 4 of \cite{norazHuntingAntisolarDifferentially2022} for a discussion). Typically, when metallicity is increased at a given stellar mass and age, the opacity also increases, and thus so too does the temperature gradient within the star. The convective zone becomes deeper in proportion to the stellar radius, with longer convective turnover times at its base because of the higher inertia induced by a higher density. This then likely modifies the convective scale distribution and also decrease the Rossby number of the star, which subsequently impacts the dynamics of the large-scale flows \citep{bessolazHUNTINGGIANTCELLS2011}. 
Such metallicity effects may lie at the origin of the observed differences in the magnetism of HD 173701 (mentioned in Section \ref{sec:stell_large_scale_obs}). \cite{karoffInfluenceMetallicityStellar2018} suggest that the higher metallicity of that star could either enhance the magnetic field generated or the observed facular contrast, yielding then the large amplitude brightness variations observed during the activity cycle. Comparisons between solar twins of different metallicities are needed to either confirm or decipher such mechanisms. To guide those observations, local-area numerical studies of the metalicity impact on photospheric convection have already started \citep{witzkeSmallscaleDynamoCool2022}, and studies using global simulations are currently being investigated \citep{noraz2022}.

A major point to mention here, is that clear detections of anti-solar rotators are still pending for solar-type stars on the main sequence. In addition to the decrease of sensitivity regarding slow-rotation for current observational techniques, recent results highlight that a change in nature of the dynamo may be induced by a rotational transition toward anti-solar rotation \citep{karakMagneticallyControlledStellar2015,vivianiStellarDynamosTransition2019,norazImpactAntisolarDifferential2022,brunPoweringStellarMagnetism2022}. In that context, a possible disappearance of star-spots is currently discussed in the community, which would make rotational characterization of the high-Rossby regime even more difficult via photometric techniques. An active search is currently underway to better constrain this regime for solar-type stars \citep{reinersDifferentialRotationStars2007,reinholdDiscriminatingSolarAntisolar2015,benomarAsteroseismicDetectionLatitudinal2018,norazHuntingAntisolarDifferentially2022}, and its implications for the solar case.

\section{Discrepancies between observations and models and possible solutions }\label{sec:3}

Significant difficulties remain when making direct comparison of the observed large-scale motions on the Sun and those produced in numerical simulations still exist.  Most fundamentally, until recently~\citep{hotta2021}, all global spherical shell models of global convection and circulation produced anti-solar differential rotation profiles (a slow equator and a fast pole) at solar rotation rates and at the solar luminosity, and no local area models of solar convection produces a peak in the photospheric horizontal velocity a supergranular scales, as observed. These discrepancies likely reflect a mismatch in convective amplitudes at depth, that has been come to be known as the `convective conundrum' \citep{2016AdSpR..58.1475O}. The global implications were first recognized when increased numerical resolution in simulations resulted in lower diffusivity, faster convective flows, and consequent difficulty in achieving a solar-like differential rotation profile~\citep{miesch2008}.

\subsection{The convection conundrum}
\label{sec:convective_conundrum}

\begin{figure}[!t]
\centering
\includegraphics[width=\textwidth]{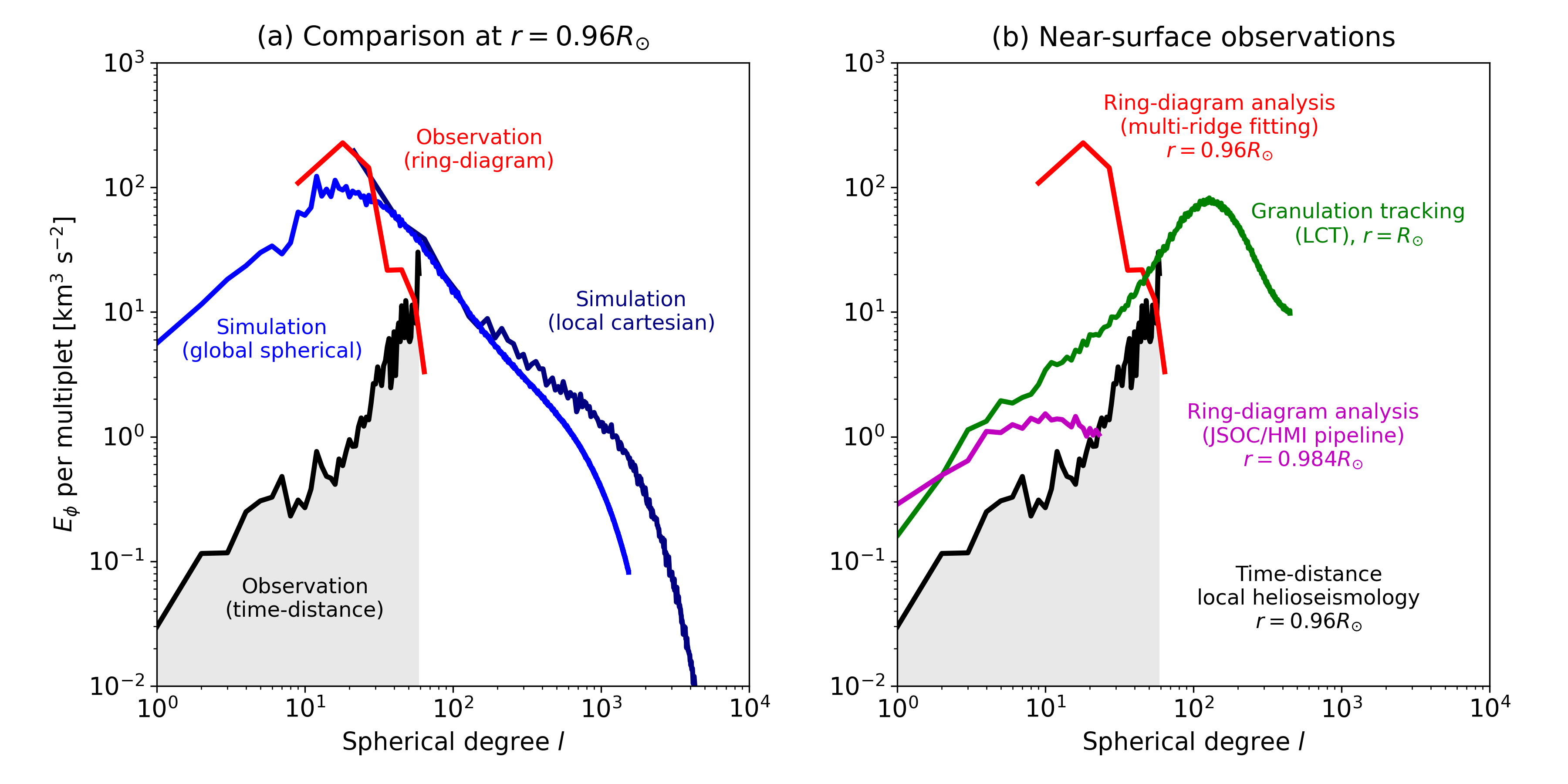}
\caption{
Horizontal velocity power spectra near the solar surface.
(a) Comparison of the spectra between numerical simulations and the observations at $r=0.96R_{\odot}$ \citep{3.DFU3SQ_2023}.
\textit{Blue}: A global full-spherical simulation of rotating magneto-convection by \citet[][]{hotta2021}.
\textit{Navy}: A local cartesian box simulation of solar convection by \citet[][]{hotta2019}.
\textit{Black (gray area)}: Observational upper limits inferred by deep-focusing time-distance helioseismic measurement of \citet{hanasoge2012}, revised recently by \citet{proxauf2020t}.
(b) Comparison of the spectra obtained by various observational measurements at various depths.
\textit{Red}: Multi-ridge fitting ring-diagram analysis by \citet[][]{greer2015}, revised recently by \citet[][]{nagashima2020}.
\textit{Green}: Local correlation tracking of surface granulation \citep{proxauf2020t}.
\textit{Magenta}: The SDO/HMI ring-diagram pipeline \citep{bogart2011a,bogart2011b,proxauf2020t}.
All observations reported in the above plots  are available online \citep{3.DFU3SQ_2023}.
}\label{fig_bekki:conundrum}
\end{figure}

There is an observation mismatch between flow velocities on the Sun and those found in numerical simulations.  In the surface layers this is dramatically illustrated by 
the absence of supergranulation in local area models, compared to its importance in photospheric observations.  That aspect of the convective conundrum is discussed in Section~1.2 and 2.2 above.  In summary, these disparities suggest reduced convective amplitudes with depth, by a factor of about 2.5 below 10 Mm, with several causes for this reduction suggested in the literature, any one or more of which would suffice.

Convective amplitudes in the near solar surface region of the solar convection zone can be measured using a variety of local helioseismic techniques. As illustrated by Figure~\ref{fig_bekki:conundrum} these do not agree with each other or with numerical simulations~\citep{hanasoge2012,gizonHelioseismologyChallengesModels2012, greer2015}. 
In general models produce flow with significant power at low wavenumbers, monotonically increasing to low wavenumber, while observations indicate reduced power there.  A modeling exception is the recent~\cite{hotta2021} simulation, in which the horizontal velocity power rolls over at large scales, though at scales somewhat beyond that of supergranulation.  An observational exception is the ring-diagram result of~\cite{greer2015} which shows significant power at scales larger than supergranulation at a depth of 0.96 $R_{\odot}$.  

While refinement and revisions of the techniques employed have brought measurements and models closer together \citep[][]{nagashima2020,proxauf2020t}, there are significant and important remaining discrepancies.  It is imperative to resolve these, not just to understand the Sun and its dynamo, but because the solar case serves as the touchstone for stellar modeling.  

\subsection{Columnar convective modes}

Numerical simulations of rotating convection have repeatedly found that the giant-cell convection tends to exist as banana cells in a strongly rotationally-constrained regime \citep[e.g.,][]{miesch2000}.
These banana cells are located outside the tangential cylinder and can be seen as north-south aligned downflow lanes across the equator.
They are known to propagate in a prograde direction with frequencies higher than the local differential rotation rate \citep[][]{miesch2008}.

The prograde propagation of banana cells can be understood in terms of a special class of inertial modes called \textit{columnar convective modes} or \textit{thermal Rossby modes} \citep[e.g.,][]{busse2002,miesch2008,bekki2022b}.
They are $z$-vorticity waves arising from the compressional $\beta$-effect due to the strong background density stratification.
A linear dispersion relation of the columnar convective modes was first derived by \citet[][]{glatzmaier1981} using a cylindrical model and later by \citet[][]{hindman2022}  in Cartesian geometry.
The most realistic spherical-shell model has recently been presented by \citet[][]{bekki2022a}.
They have shown that the dispersion relation of the columnar convective modes is very sensitive to the superadiabaticity $\delta$, and that the modes become convectively-unstable (exponentially growing) when the background is slightly superadiabatic ($\delta>0$).
These modes have the associated Reynolds stress $\langle v_{r} v_{\phi}\rangle>0$ and $\langle v_{\theta} v_{\phi}\rangle>0 \ (<0)$ in the northern (southern) hemisphere, implying that they transport the angular momentum radially upward and equatorward outside the tangential cylinder.
The great importance of \textit{banana cells} on the establishment of the solar-like differential rotation has been repeatedly appreciated in previous literature \citep[][]{kapyla2011,gastine2013,hotta2015,matilsky2020,camisassa2022}.

Despite their significant importance, columnar convective modes have never been observed on the surface of the Sun, though very large-scale flows have been deduced from correlation tracking of the solar supergranulation~\citep[][]{2013Sci...342.1217H, 2021ApJ...908..160H}
and possible indirect evidence for these flows comes from measured alignment of the solar supergranulation~\citep[][]{2004ApJ...608.1167L, 2011ApJ...726L..17N}.
It remains a mystery why the columnar convective modes are not directly detected in the solar surface observations.
One possible scenario is that they indeed exist hidden in the deep convection zone with substantial amplitudes but are concealed by surface small-scale convective feature \citep[][]{guerreroDIFFERENTIALROTATIONSOLARLIKE2013}.
The other scenario is that they are simply absent in the Sun or too weak to be detected~\cite[][Section 2.2 above]{1986ApJ...304..828V, lord2014}.
If the latter is the case, the angular momentum needs to be transported by something other than the large-scale columnar convective modes.
For instance, the equatorward angular momentum transport can be achieved in the Sun by the Reynolds and/or Maxwell stresses at much smaller spatial scales.

\subsection{Differential rotation}
\label{sec:hotta_diffential_rotation}

An important aspect of the convective conundrum is its implications for differential rotation. In early phase of the solar differential rotation research, the solar differential rotation profile could be reproduced relatively easily. As supercomputer power grew, allowing simulation of higher resolution, the problem became apparent. Simulations began to fail to reproduce the solar-like differential rotation profile at solar rotation rates. \par

A well-known feature of differential rotation is that a fast equator (poles) is obtained with strong (weak) rotational influence \citep[e.g.,][]{gastine2013,featherstone2015,karakMagneticallyControlledStellar2015}. The rotational influence is measured by the Rossby number $\mathrm{Ro}=v/(2\Omega H)$, where $v$ is a characteristic convective velocity and $H$ a characteristic spatial scale. Although we believed that the Sun is in a low Rossby-number regime, high resolution simulations most often produce anti-solar differential rotation profiles (fast poles).  High resolution simulations introduce the small-scale turbulence which is important for heat transport but is only very weakly rotationally constrained.  This has been recently confirmed with scale-dependence analyses of the angular momentum flux \citep{mori_2023MNRAS.519.3091M}. The small-scale turbulence tends to transport the angular momentum radially inward which leads one-cell meridional flow, which transports angular momentum and accelerate the poles \cite[see also][]{featherstone2015}. In other words, high-resolution decreases the convective spatial scale $H$ placing the simulation in a high Rossby number regime, resulting an anti-solar differential rotation.\par
Three mechanisms have been employed in global simulations in order to maintain a solar-like differential rotation profile, i.e, to maintain the low Rossby number, in the face of vigorous smaller-scale convective flows.
\begin{itemize}
    \item Increasing the rotation rate $\Omega_0$ \citep{brownRapidlyRotatingSuns2008,nelson_2013ApJ...762...73N,hotta_2018ApJ...860L..24H}.
    \item Reducing the luminosity $L$ \citep{hotta2015}, which leads to a reduction of $v$.
    \item Increasing viscosity $\nu$ and/or thermal conductivity $\kappa$ \citep[][]{miesch2000,miesch2008,fan_2014ApJ...789...35F,hotta_2016Sci...351.1427H}, which leads to a reduction of $v$ and an increase of $L$.
    \item Increasing radiative conductivity $\kappa_{\rm rad}$ \citep[][]{kapylaEffectsSubadiabaticLayer2019,noraz2022}, which leads to a reduction of $v$ while conserving $L$.
\end{itemize}
None of these reflect a possible physical mechanism operating on the Sun but not captured in simulations. They cannot provide an answer to the question: Why does the Sun have such a low Rossby number?\par
Recently, \cite{hotta2021} reproduced the solar-like differential rotation in extremely high resolution simulation without using these manipulations. \cite{hotta2022} showed that in that simulation the strong magnetic field maintained by the convection when the magnetic diffusivity is very low, as it is in their simulation, causes angular momentum to be transported outward. A double-cell meridional flow is generated and that flow transports angular momentum equatorward, resulting in the solar-like differential rotation profile achieved.  An important point is that the \cite{hotta2021} simulation convection remains in a high Rossby number regime but the  columnar convective modes no longer play an dominant role in the angular momentum transport. The Rossby number range in which a solar-like differential rotation regime is possible is under investigation, and recent studies suggest that this range depends sensitively on the Prandtl number  \citep[][]{kapylaTransitionAntisolarSolarlike2022,noraz2022}.

\section{Future prospects}
\label{sec:fugure_prospect}

From a theoretical view point, some more significant progress can be made in the near future. \cite{hotta2021} show that the high-resolution simulations remain a promising approach to understand the solar large-scale flows. Simulations have not yet reached {\it numerical convergence}, where by numerical convergence we mean that when we double the resolution the large-scale structure does not change. The solar turbulent convection in the simulations has a spatial spectrum which has more power at low wavenumbers than is observed on the Sun. With the injection scale at 200 Mm and the dissipation scale around 1 cm, it is impossible to directly resolve all the energy containing scales with numerical simulations. 
Although higher-resolution numerical simulations are required to better understand the nature of turbulent convection, it is also crucial to construct reliable turbulence models which can mimic the essential features of unresolved flows.

Such models require a deep understanding of the key physical components of the convection.  
Important progress has been made in understanding the essential role of rotation~\citep[e.g.,][and references therein]{2021PNAS..11822518V} and preliminary work has begun to characterize the highly nonlocal convective flows that result from radiative cooling of the photosphere which may play an important role in heat transport and allow a mean gradient much closer to isentropic than that achieved by current simulations~\citep{2016ApJ...832....6B, 2016ApJ...829L..17C}. Moreover, radiative heating of the lower convection zone likely plays an important role, one that has only begun to be examined.
 \cite{brun_2011ApJ...742...79B} have undertaken an important study of the interaction between the convection and radiation zone, but, as \cite{kapyla_2019A&A...631A.122K} pointed out, a fully consistent treatment of the overshoot layer requires huge numerical resources, which are currently not available \citep[see also][]{hotta_2017ApJ...843...52H}.
 Solving these problems on convection is essential to understanding the large-scale flow dynamics.
 
Another direct approach is to combine in one simulation the radiative magnetohydrodynamics of the photosphere and the global scale convective motions.  Including the photosphere may have a significant impact on the deep large-scale convection~\citep{spruit_1997MmSAI..68..397S}.  \cite{hotta2019} found only weak (or no) influence from the photosphere on the deep convection, however their result may not reflect a lack of importance of these flows, but the difficulties faced in maintaining them with depth given the diffusivities required in global models. Even when the resolution of the simulation is increased (and numerical or explicit diffusivities are reduced), the problem persists because the horizontal scales of the flow structures are also decreased.  What is required to maintain nonlocal transport in a simulation is that the diffusion time across a downflowing plume be greater than the transit time of that fluid across the fluid layer.  This is extremely difficult to achieve.  Because of the steep stratification, local area radiative magnetohydrodynamic simulations, and ambitious global models which include an upper radiative boundary, may be able to produce granular scale downflow structures but not the low thermal diffusivity required for them to maintain their role in transport at depth.

Over the past fifty years, tremendous progress has been made in understanding and simulating solar convection and the global scale flows that result in a rotating domain.  The Sun allows us to directly confront that progress with observations that, with the advent and development of helioseismology over the same time period, have also made previously unimaginable advances. New diagnostics based on the study of the solar inertial modes are expected to provide additional constraints on the physics of the deep convection zone. Some gaps in the observations and some fundamental issues in our understanding however remain. Resolving those will not only advance our understanding of the origin of large-scale global flows, but will also allow us to more robustly model the solar dynamo, perhaps predict solar behavior critical to human activities in space and on Earth, and extend our understanding to other stars for which comparison with data will remain less constraining. Understanding the Sun is thus a touchstone activity.

\bmhead{Acknowledgments}
We acknowledge the support from ISSI Bern for our participation in the workshop. 
We thank Zhi-Chao Liang for helpful comments on the manuscript.  
HH is supported by JSPS KAKENHI grant Nos. JP20K14510, JP21H04492, JP21H01124, JP21H04497, and MEXT as a Program for Promoting Researches on the Supercomputer Fugaku (“Toward a unified view of the universe: from large-scale structures to planets,” grant No. 20351188).
YB, LG and QN acknowledge financial support from ERC Synergy Grant WHOLE SUN  810218. QN was funded in part by an INSU/PNST grant and CNES (Solar Orbiter).  MPR acknowledges partial support for this work by the National Science Foundation under award number NSF 1841100.

\section*{Ethics Declarations}
{\bf Competing interests}
\medskip
The authors declare they have no conflicts of interest.

\bigskip








\bibliography{reference}

\end{document}